\documentclass[]{aa}
\usepackage{graphicx}
\usepackage{amssymb,amsmath}
\usepackage{threeparttable}
\usepackage{txfonts}
\usepackage{float}
\usepackage{subfigure}
\usepackage{tabularx}
\usepackage{natbib}
\usepackage{epsfig}

\begin{document}

\title{The Starburst-AGN connection: quenching the fire and feeding the beast}

\author{Jorge Melnick\inst{1}$^,$\inst{2}
\and Eduardo Telles\inst{2} 
\and Roberto De Propris\inst{3}
\and Chu Zhang-Hu\inst{4}}

\institute{European Southern Observatory, Av. Alonso de Cordova 3107, Santiago, Chile
    \and
 Observatorio Nacional, Rua Jos\'e Cristino 77, 20921-400 Rio de Janeiro, Brasil
    \and
Finnish Centre for Astronomy with ESO, University of Turku, V{\"a}is{\"a}l{\"a}ntie 20,    
   Piikki{\"o}, 21500, Finland
   \and
 Department of Astronomy, Nanjing University, Nanjing 210093, P. R. China}
\date{}
\offprints{Jorge Melnick, \email{jmelnick@eso.org}}

\maketitle

\abstract{The merger of two spiral galaxies is believed to be one of the main channels for the production of  
elliptical and early-type galaxies. In the process, the system becomes an (ultra) luminous infrared 
galaxy, or (U)LIRG, that morphs to a quasar, to a K+A galaxy, and finally to an early-type galaxy. 
The time scales for this metamorphosis are only loosely constrained by observations. In particular, 
the K+A phase should follow immediately after the QSO phase during which the dust and gas 
remaining from the (U)LIRG phase are expelled by the AGN. An intermediate class of QSOs with K+A 
spectral signatures, the post-starburst QSOs or PSQ, may represent the transitional phase between 
QSOs and K+As. We have compiled a sample of 72 {bona fide} $z<0.5$ PSQ from the SDSS DR7 
QSO catalogue. We find the intermediate age populations in this sample  to be on average significantly
weaker and metal poorer than their putative descendants, the K+A galaxies. The typical spectral 
energy distribution of PSQ is well fitted by three components: starlight; an obscured power-law; and a 
hot dust component required to reproduce the mid-IR fluxes.  From the slope and bolometric luminosity of 
the power-law component we estimate typical masses and accretion rates of the AGN, but we find little
evidence of powerful radio-loud or strong X-ray emitters in our sample. This may indicate that the 
power-law component originates in a nuclear starburst rather than in an AGN, as expected if the bulk 
of their young stars are still being formed, or that the AGN is still heavily enshrouded in dust and gas. 
We find that both alternatives are problematic and that more and better optical, X-ray, and mm-wave
observations are needed to elucidate the evolutionary history of PSQ.}


\section{Introduction}
\label{intro}

The possibility that at least some elliptical galaxies could be the end result of the merger of two spirals of similar masses was 
first proposed by \cite{Toomre72} (see also \citealt{Toomre77}) and has since been explored in numerous investigations (see e.g. \citealt{Hopkins08a} 
for a recent review).  While intriguing, the idea of the metamorphosis of spirals into ellipticals has been questioned on dynamical 
and photometric grounds. Nevertheless, one by one these objections have been resolved by increasingly sophisticated observations and numerical 
modelling. 

The main difficulty lies in suppressing star formation in the merger of two gas-rich systems to return a quiescent remnant. This process
(quenching) may be aided by feedback processes, and especially by the activation of an active nucleus (expected if mergers drive gas
and dust to the centre by tidal torques (e.g. \citealt{Mihos96}). In the detailed simulations by \cite{Hopkins08a,Hopkins08b} the merger of two
gas-rich systems produces an AGN that eventually stops or prevents further star formation. The resulting remnant goes through a brief
AGN phase and is then identifiable as a post-starburst (K+A) galaxy as the initial star formation episode fades away.  A recent investigation 
has revealed that a median 50\% of the stellar mass in a sample of 808 K+A  galaxies has intermediate ages and high metallicities 
\citep{Melnick2014}, consistent in fact with the metallicities of elliptical galaxies of similar masses.  Thus, the present day stellar populations 
of at least some elliptical galaxies appear to have been formed in a colosal burst of star formation induced by the merger of two gas-rich galaxies
of similar masses (the preferred model for the origin of K+A galaxies \citealt{Snyder11}).

The signatures of this process may be difficult to identify. There are several lines of evidence now favouring a secular scenario for AGN evolution (e.g. \citealt{Jahnke11}). X-ray selected AGN hosts in deep HST imaging are not observed to be more disturbed than a similar sample of inactive galaxies at the same redshifts (e.g. \citealt{Cisternas11,Kocevski12,Villforth14} and references therein), while both \cite{DePropris14a} and \cite{Scott14} do not find evidence for a strong enhancement of the merging fraction in closely interacting pairs, despite the presence of tidal signatures. Similarly, \cite{Teng12} searched for AGN activity in companion galaxies to QSO hosts and detected just one AGN in their sample of 12 galaxies. 

It is of course possible that early QSOs are highly obscured: \cite{Koss11} shows that hard X-ray selected AGNs, which are often Compton thick, are hosted by highly disturbed galaxies, while \cite{Urrutia10} find that a large fraction of Type II (red, dust obscured) QSOs lie in morphologically disturbed systems. \cite{Canalizo2013} argue that QSOs with strong absorption lines have evidence of tidal interaction, but this is only present at low surface brightness levels, which argues for gas-poor, low mass, or even old remnants, that may or may not be connected with the observed present-day activity.

If we accept the QSO-merger model, we expect  that at least some young QSOs may be associated with a post-starburst signature as they have just suppressed star formation and are emerging from their dust-enshrouded phase. In a previous paper \citep{DePropris14b}  we have shown that no K+A galaxy at any age hosts an AGN, although there is clear evidence of feedback (see also \citealt{Wong12} for a similar conclusion). Therefore, suppression of star formation and QSO activity must rapidly follow each other (see \citealt{Teng12}). Thus, the particular  class of quasars known as post-starburst quasars (PSQ; see \citealt{Cales2013} for a recent review), would be the immediate ancestors of K+A  galaxies and therefore the descendants of recent mergers.

Observationally, PSQ are characterised by strong emission lines and strong absorption of the high Balmer lines. \cite{Cales2013} concluded that PSQ with early-type hosts are likely the result of major mergers, so this class of objects may offer a way to witness the final stages of the metamorphosis of spirals into early-type galaxies. In this paper we present a systemic study of the properties of a sample of 72 PSQ from the SDSS selected from the DR7 Catalog of Quasars \citep{Schneider2010}, paralleling our study of K+A  galaxies. We use the SDSS spectra to derive the stellar populations, which are then used to predict the broad-band SEDs from the far ultraviolet (FUV) to the 
mid infrared (mid-IR(. Throughout this work we adopt the following cosmological parameters: $\Omega_M=0.3$, $\Omega_{\Lambda}=0.7$, and $H_0=70$km s$^{-1}$ Mpc$^{-1}$.

\section{Data and models}

The DR7 Quasar Catalog (DR7Q) contains data for 105,783 SDSS quasars selected as described by \cite{Schneider2010}. We downloaded the SDSS spectra of 
the 8492 objects in the DR7Q with $z\leq0.5$, which we searched for signatures of post-starburst stellar populations using a slightly more relaxed criterion 
than that used by \cite{Goto2006}: we looked for objects with (rest frame) equivalent withs of H$\delta$ greater than $\rm EW({H\delta})>3$\AA\ in absorption. We found that about 1.5\% of the objects (125 out of 8492) satisfied this criterion. 

{ Our PSQ fraction is significantly lower than the fraction of 4.2\% PSQs found by \cite{Goto2006} in a volume limited sample of SDSS galaxies. This is not surprising given that Goto gives the fraction of all AGN that have strong $H_{\gamma}$ in absorption, while our sample is restricted to objects classified as QSOs, that is, only to AGN with broad emission lines. In fact, the three examples shown by  \cite{Goto2006} are classified as (Seyfert~2) galaxies in the SDSS and therefore are not included in the DR7Q.}


We visually inspected the spectra for spurious detections, which led to a further reduction of the sample to 83 objects; while some were clearly 
not PSQs, the S/N ratios of others was not sufficient to ascertain the presence of strong $H\delta$ in absorption.  Finally we ran the {\sc Starlight} code \citep{Cid2005} 
to model the stellar populations of these 83 objects using exactly the same prescriptions of \cite{Melnick2014} for the K+A  galaxies: BC03 \citep{BC03} stellar library
for 25 ages and 5 metallicities. From the {\sc Starlight} fits we found that the SDSS spectra of 10 out of our 83 bona fide PSQs were of insufficient quality to derive
reliable stellar populations, either because the S/N was too low ($S/N<2$ at 4100\AA), or due to significant gaps at some critical wavelengths. 

{ The essential statistical properties of our final sample of 72 PSQ are presented in Figure~\ref{sample} in comparison the properties of the K+A galaxies from \cite{Melnick2013}. The comparison is relevant because the main purpose of our investigation is to establish whether PSQ can be the progenitors of K+As. 

\begin{figure*}
  \hspace{-0cm}\includegraphics[height=6.0cm]{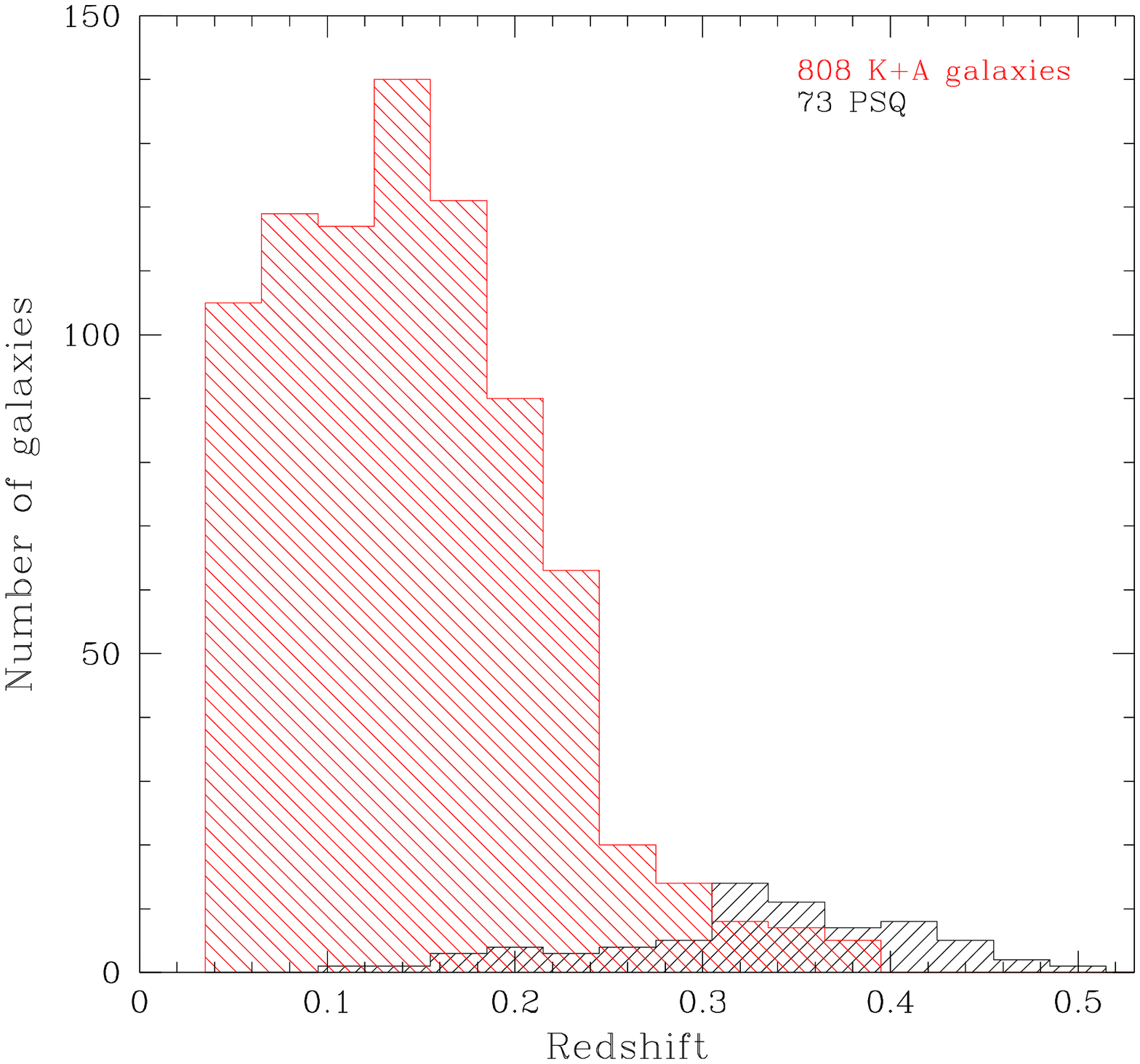}\includegraphics[height=6.0cm]{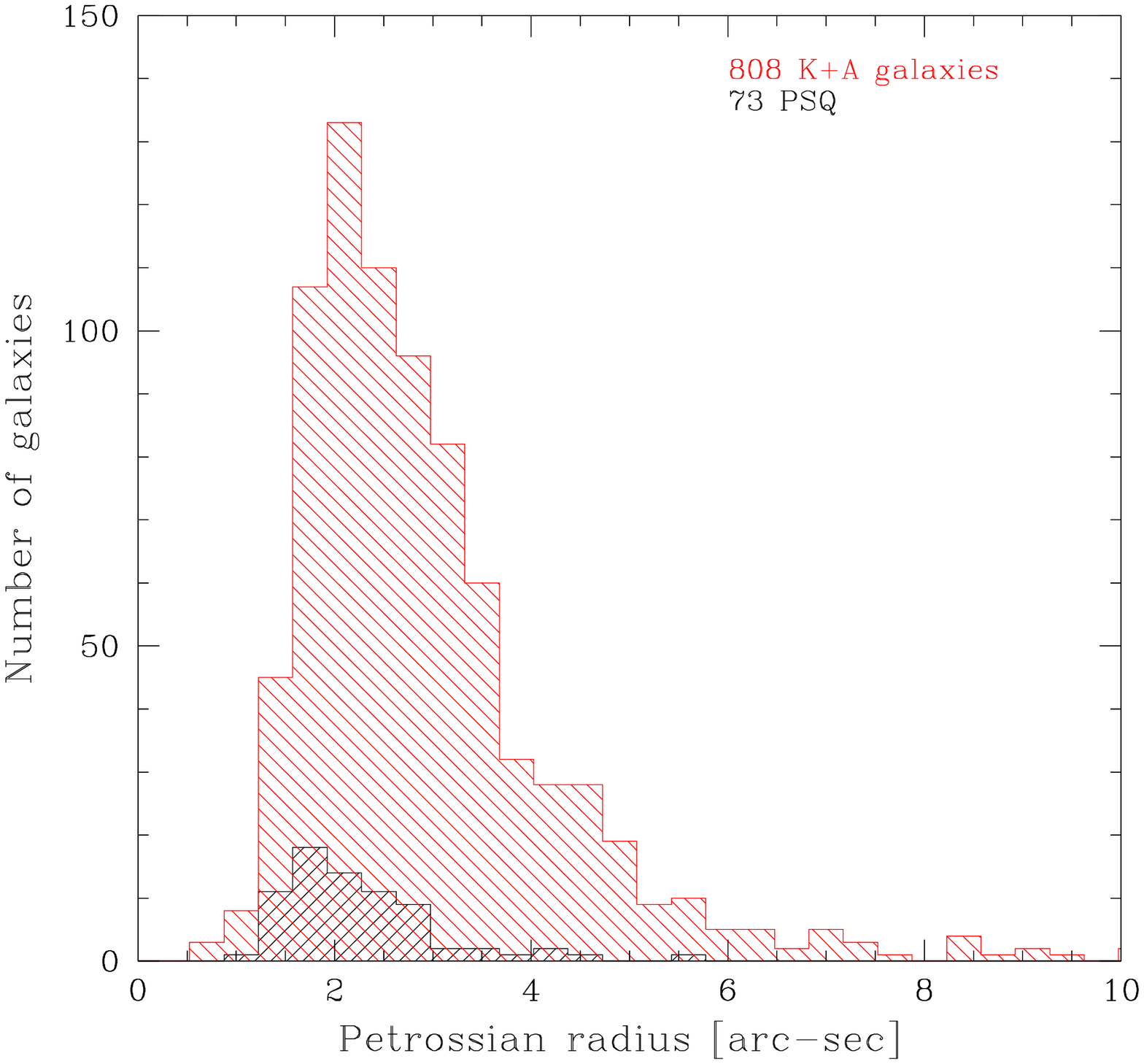}\includegraphics[height=6.0cm]{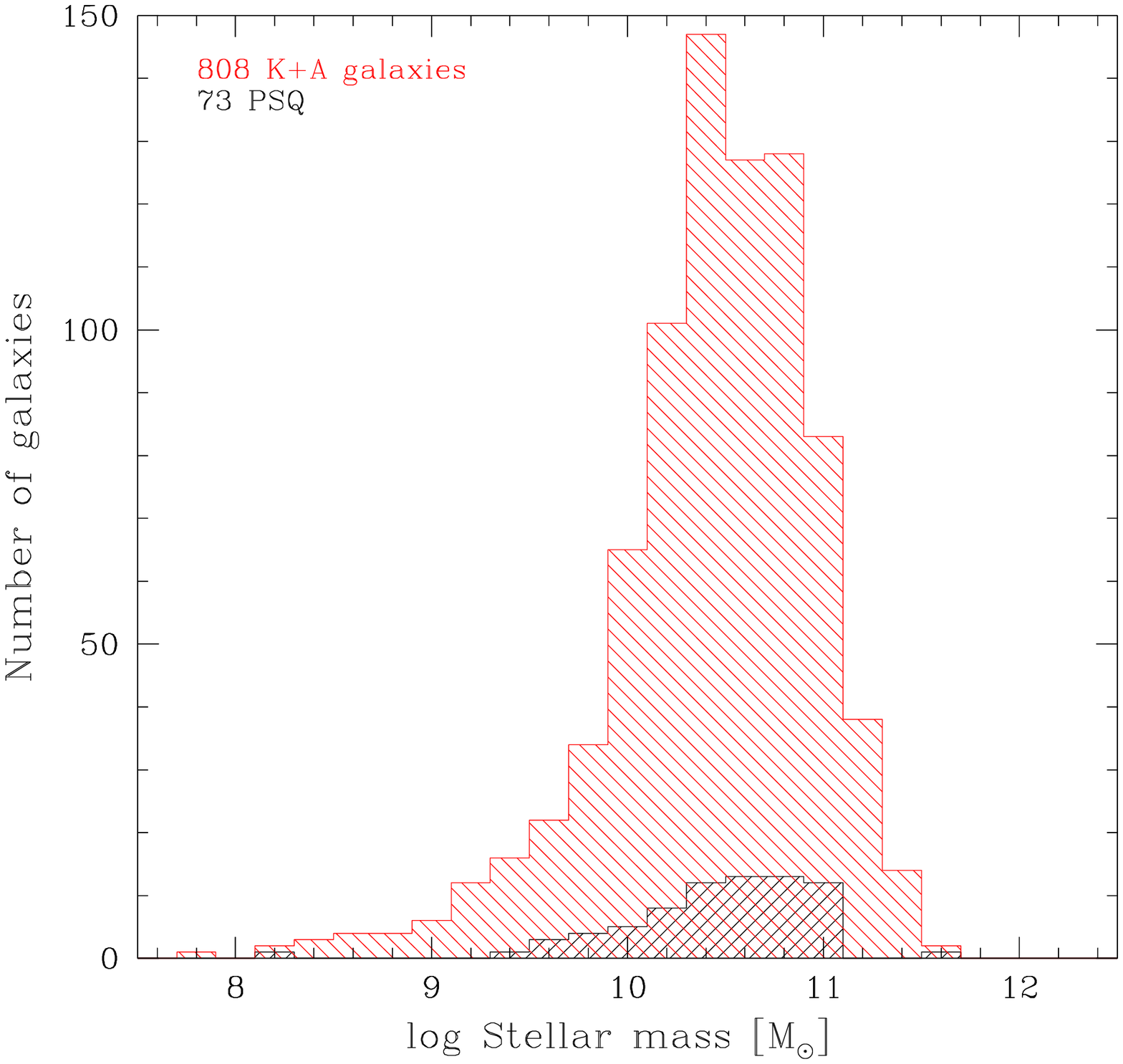}
  \caption {Statistical properties of the PSQ sample (in black) compared to those of their putative descendants the K+A galaxies (in red). The number of objects in each bin has been divided by ten for the K+A radii and by five for the masses.'}
\label{sample}
\end{figure*}

Figure~\ref{sample} shows that - as QSOs - PSQ are rare in the local Universe, even compared to K+A galaxies. On the other hand, the stellar masses of PSQ appear to be comparable to, but somewhat larger than, K+As. We note, however, that the PSQ masses are rather uncertain given the generally low S/N of our spectra.  More surprising is that the Petrossian radii of the two samples are comparable given the differences in redshift. While this is probably due to the relatively low resolution of the SDSS images, it confirms that, as discussed by \cite{Melnick2013}, the photometry and spectroscopy of our PSQ sample are not significantly affected by systematic aperture matching effects.
}

The star formation histories for our 72 { bona fide} PSQs were combined to build the spectral energy distributions (SEDs) of the stellar components from the  FUV to the mid-IR using the BC03 stellar library to convert masses and ages to photometric fluxes, which we compared to the photometric observations.  { We emphasise that we do not use the spectral fluxes to construct the SEDs, but the stellar populations derived from the spectra. This allows us to essentially  {\em extend} the spectral coverage to the FUV and mid-IR.}

The essential characteristics of the broad-band photometric dataset are: optical (ugriz) from the SDSS \citep{Abazajian2009}; Vacuum UV data are from the GALEX imaging databases  \citep{Morrissey2007}; Near-infrared data in JHK come from 2MASS  \citep{Skrutskie2006}; and Mid-infrared data from the WISE \citep{Wright2009} datasets. The photometry and the input SDSS spectra were corrected for Galactic foreground extinction using the values of given in the DR7Q \citep{Schneider2010} using the standard Galactic reddening law (\citealt{Cardelli1989}; henceforth CCM). 


The synthetic spectral energy distributions are calculated as

\begin{equation}
F(\lambda)= k {\sum_{i=1}^{25}}{\sum_{j=1}^5} S_{ij} M_{ij} (\lambda),
\end{equation}
\smallskip\\
{where $M_{ij}$ are the  BC03 models} for the $i=1,...,25$ age bins and $j=1,..,5$
metallicity bins, and $S_{ij}$ are the mass fractions returned by {\sc Starlight} for the relevant age and
metallicity bins.  The photometric zero-point $k$ is independent of wavelength.  

\section{Results}
\subsection{Stellar Populations}
\label{popse}

\begin{figure*}
  \hspace{-0.5cm}\includegraphics[height=7.0cm]{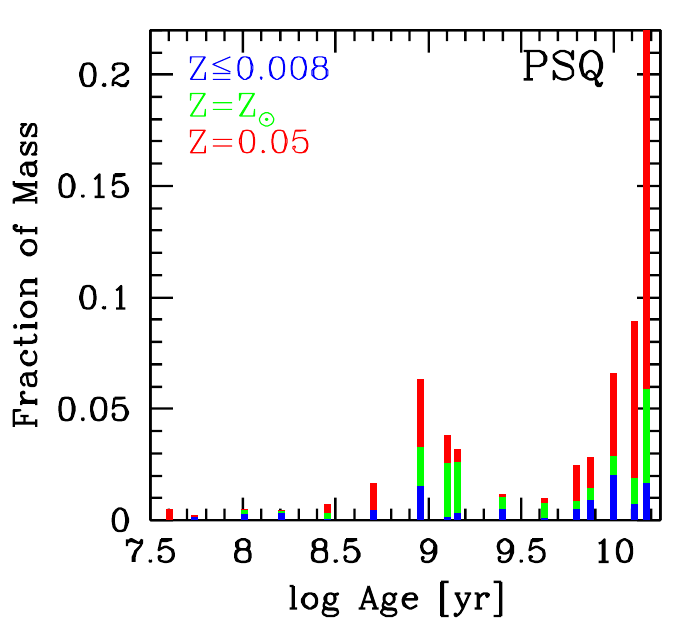}\hspace{1cm}\includegraphics[height=7.0cm]{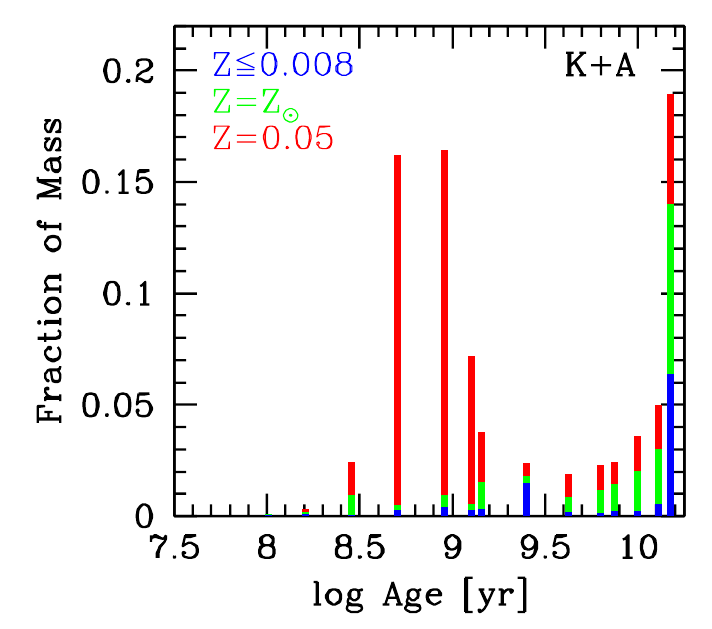}
  \caption {Comparison between the stellar populations of PSQ on the left and of K+A  galaxies on the right for three broad metallicity bins.}
 \label{popos}
\end{figure*}

Figure~\ref{popos} presents in the left panel the average stellar populations of the full sample of 72 PSQ analysed with {\sc Starlight}. These plots give fractions of 
the total stellar mass as a function of age for three broad metallicity bins (metal poor; solar; over-solar) chosen to make the plot less cluttered.  For comparison, 
the right panel presents the mean stellar populations of our sample of 808 K+A  galaxies from \cite{Melnick2014}.  Two remarkable differences are immediately obvious: (1) the average fraction of mass in intermediate age stars is about 15\% in PSQ compared to $\sim45$\% for K+A  galaxies; and (2), while in K+As the vast majority of the intermediate-age stars are metal 
rich, in PSQ this is the case for only a small fraction of the stars. 

{ It is notoriously difficult to estimate formal statistical errors for the stellar populations derived from population synthesis modelling (see e.g. \citealt{Melnick2014} for a discussion in the case of K+A galaxies and {\sc Starlight}). Systematics dominate over formal fitting errors, in particular the well known degeneracies between age and metallicity. In the case of PSQ one should be particularly concerned by the effect of a featureless power-law continuum from the AGN. 

{{\sc Starlight} offers various possibilities to include a power-law continuum, and in particular, to deal with different extinctions for different populations. We ran many experiments to establish whether a combination of power-law and extinction could affect the resulting stellar populations. We found that, while including these combinations may change the detailed distribution of ages and metallicities, the total fractions of intermediate age stars remained substantially unchanged. 

We obtained very consistent results by either including a power-law component of slope $\alpha=-1.5$ ($f_{\lambda}\propto(\lambda/0.55\mu)^{-1.5}$) reddened by $A_V=1.3$, as determined from our SED fitting discussed in the next section, or by including an un-reddened power-law of the same slope and allowing {\sc Starlight} to choose the best fitting extinction for each object. The former is shown in Figure~\ref{popos}, and is the one we used to compute the total stellar masses, but the difference between the two methods are surprisingly small.

From these and the other experiments we ran, we are confident that the differences between the stellar populations of PSQ and K+As shown in Figure~\ref{popos} are not due to our choice of how to model the power-law continuum.}

Another systematic effect is contamination of the Balmer stellar lines by nebular emission, which could result in lower fractions of intermediate-age stars. Visual inspection of the spectra shows this effect to be present in some of our objects, but even at H$\delta$ (we mask H$\alpha,\beta,\gamma$) the contamination is very small. In order to quantify this effect we subtracted the population synthesis model from the original spectra to generate a pure nebular spectrum. We then subtracted the smoothed nebular spectrum from the observed spectrum to generate an emission-line free spectrum, which we fitted again with {\sc Starlight}.

This procedure inevitably adds noise to our already rather noisy SDSS spectra. We have tested the procedure with rather discouraging results: we get no net increase in the fraction of intermediate-mass stars and the metallicities do not change significantly. Masking out H$\delta$ in the original spectra gave similar negative results. We also tested stacking the spectra of all the objects that allows us to remove the emission lines from the stacked spectrum without adding noise. The change in the stellar populations with and without removing the emission lines was negligible and EW(H$\delta$) changed by less than 10\%. 
 
We performed a number of other tests to evaluate the systematical errors in our results:  changing spectral range to eliminate the noisiest spectral regions; changing the reddening law; and masking various spectral features. While the stellar populations change in detail, the overall fractions and metallicities of the intermediate-age component never changed significantly and in particular never increased. We are confident, therefore, that the discrepancy between PSQ and K+As is not due to errors in the fitting procedure.}

\subsection{Synthetic SEDs}

\begin{figure*}
\includegraphics[height=8cm]{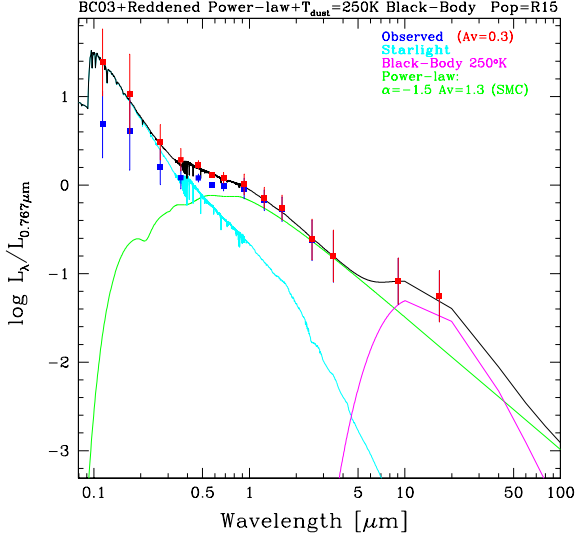}
\caption {Best fit model SED for the average of the 72 K+A  galaxies in our sample compared to the average observed broad-band fluxes corrected for extinction as described in the text and normalised to the SDSS $i'$-band ($F_{0.767\mu m}$).  The blue symbols show the observed photometric data points and the red symbols show the observations corrected for extinction using the visual extinction shown in parenthesis in the figure legend, and the GD1 extinction law. The three components of the fit are shown colour-coded as follows: cyan shows the contribution of the stars computed from the {\sc Starlight} population-synthesis fits to the SDSS spectra; green shows the obscured power-law; and the magenta represents the $T_{dust}=250K$ hot dust component, which is required by the mid-IR data. The sum of these three components is shown in black. The error bars {\em are not observational errors} but correspond to the $1\sigma$ dispersions of the normalised fluxes for all the objects included in each band, the number of which varies from 45 objects for the FUV band to 72 for the SDSS bands. The reddened power-law component was included in the population synthesis models, thus the label Pop=R15 in the figure legend.}
 \label{seds}
\end{figure*}

Figure~\ref{seds} compares the average synthetic SED computed using Equation~1 for the mean of the {\sc Starlight} stellar populations (shown in Figure~\ref{popos})  
to the observations. In order to remove the intrinsic variance in luminosity we normalised the fluxes to the SDSS $i'$-band ($F_{0.767\mu m}$). The observed fluxes show a clear excess above $\lambda\sim0.5\mu$, which is very different from the case of K+As where the excess is only observed in the mid-IR bands at $\lambda>5\mu$.  In order to fit the PSQ, therefore, in addition to a hot-dust component parametrised as a black-body of temperature $T_{dust}$ as we did for K+A  galaxies, we added a reddened power-law component as expected for AGN.

{ We emphasise that the error bars plotted in the figure are not observational photometric errors, but the rms scatter of the normalised fluxes about the mean of the sample. The mean observational errors range from 0.26 magnitudes at $22\mu$ and 0.23~mag in FUV to $<0.03$~mag in the SDSS $g'$ to $z'$ bands. Thus, while PSQ form a remarkably homogeneous class of objects at optical wavelengths, this is not the case in the UV and IR bands. This is different from K+A galaxies that have remarkably similar SEDs in the optical and in the IR up to $4.5\mu$ \citep{Melnick2014}.}

We thus need six free parameters to fit the observed SEDs:   $T_{dust}$;  $\alpha_{UV}$ -- the slope of the power-law component; and $A^{PL}_V$ -- the visual extinction of the power-law, plus the photometric zero-points of the three components: stellar populations; obscured power-law; and hot-dust. An additional free parameter is the reddening law for the stellar and power-law components, which is very relevant for the UV bands. 

In principle the extinction is a free parameter only for the power-law component. In the case of K+A  galaxies we corrected the photometric observations using 
the extinction values from the {\sc Starlight} fits. However, this is not possible, or at least is not straightforward, in the case of PSQ because the power-law component contributes a 
significant fraction of the continuum flux in the SDSS spectral band. We verified that this does not change the stellar populations significantly, but it certainly affects
the resulting extinction, which in turn affect the SED fits and thus the optical power-law continuum. We tested an iterative procedure, which is very time consuming and does not easily converge.

We therefore followed a different procedure to estimate the extinction corrections for the stellar component, $A^*_V$. First we fixed the photometric zero points of the power-law and hot-dust components by forcing each to fit the WISE $4.5\mu$ and $12\mu$ bands respectively; the power-law completely dominates the $4.5\mu$ flux and the Black-Body completely dominates the $12\mu$ emission, with negligible extinction corrections. For the stellar component we fixed the zero point in the SDSS $i-$band where stars still contribute a significant fraction of the total flux while, not being negligible like in the mid-IR, the extinction corrections are still much lower than in the bluer bands.  

After fixing the photometric zero points we fitted remaining two free parameters (we fixed the dust temperature to $T_{dust}=250$K {\it ab initio}): the slope ($\alpha_{UV}$) and the extinction $A^{PL}_V$ of the obscured power-law component, and we computed the stellar extinction $A^*_V$ by forcing the FUV flux to match the observed values (even for a small amount of extinction the contribution of the obscured power-law is negligible in the FUV).  We then recomputed the zero points on the extinction-corrected fluxes and iterated until the zero point of the stellar component converged to better than 1\%, which takes only a few iterations.

Following our work in Paper~I, we experimented with two extreme extinction laws: the canonical Galactic law of \cite{Cardelli1989} -- CCM, and the UV-strong SMC/30Dor law of \cite{Gordon2003}, which we refer to as GD1 following the nomenclature of {\sc Starlight}.  Both laws are reasonably similar in the optical and near-UV, but differ significantly in the FUV. Also, while CCM shows the strong 2200\AA\ bump characteristic of the Galaxy, the bump is absent in GD1. We verified that GD1 provides a much better fit to our PSQ data although, within the redshift range of our sample, the 2200\AA\ bump does not affect the observations.  The photometric data uncorrected for extinction are shown as blue squares in Figure~\ref{seds} while the extinction-corrected data are shown in red, for the best fit mean stellar extinction of $A^*_V=0.27$ and the GD1 reddening law. 

Figure~\ref{seds} shows that our simple three component model provides a remarkably good fit to the observations over more than two decades in wavelength from the 
FUV to the mid-IR. There is clearly some degeneracy in the parameters, notably between in $\alpha_{UV}$ and $A^{PL}_V$, although we find solutions only for $-1.2>\alpha_{UV}>-1.8$
and $A^{PL}_V>0.8$. From a visual inspection of the $\chi^2$ contours the best fit is obtained for $\alpha_{UV}=-1.5\pm0.2$ and $A^{PL}_V=1.3\pm0.3$.  Interestingly, the power-law slope is constrained to values remarkably similar to the average value observed for X-ray luminous AGN ($f_{\lambda}\propto\lambda^{-1.6}$; \citealt{Wu2012}).

{ This reddened power-law component is the one we included in the {\sc Starlight} population synthesis models described in Section~\ref{popse}.}

\section{Emission-lines}

\subsection{Gas excitation}

\begin{figure*}
\includegraphics[height=8cm]{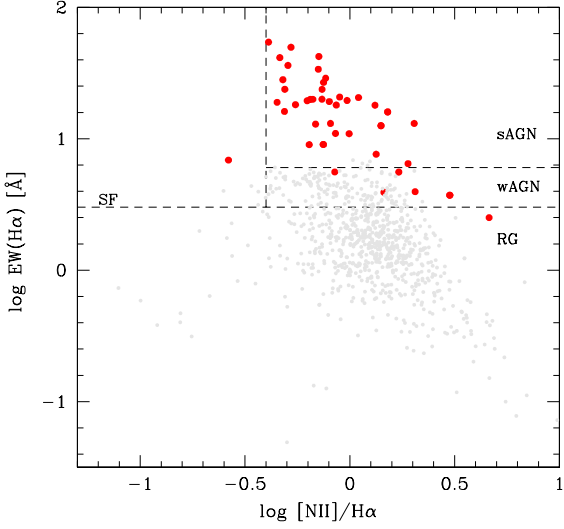}
\caption {WHAN diagram for the 72 PSQ in our sample. The grey points show our sample of 808 K+A  galaxies for comparison. }
 \label{bpt}
\end{figure*}

Figure~\ref{bpt} shows the WHAN diagram described by \cite{Cid2010,Cid2011} as a powerful emission-line diagnostic method for objects with weak emission-lines, or systems, 
such as ours, where the higher Balmer lines are severely affected by underlying stellar absorption making diagnostics such as [OIII]/H$\beta$ (or [OII]/[OIII] that requires 
extinction corrections) rather uncertain. Since most of the objects in our sample have broad H$\alpha$ emission lines, we used the package PAN in IDL \citep{Dimeo2005} 
to simultaneously fit the narrow (H$\alpha$ and [NII]) and the broad components. This was only possible for a subset of all our objects because at their redshifts the SDSS spectra around H$\alpha$ are rather noisy due to contamination from atmospheric features, and in some cases outside the spectral range.

As expected (the objects were selected from a QSO catalogue) most of the objects in our sample for which we could measure the parameters fall in the region of 
strong (and presumably real) AGN (sAGN). A few are weak AGN (wAGN formerly LINERS), while only one PSQ appears to be powered by young stars (SF).  

\subsection{Gas kinematics}

\begin{figure*}
\includegraphics[height=7cm]{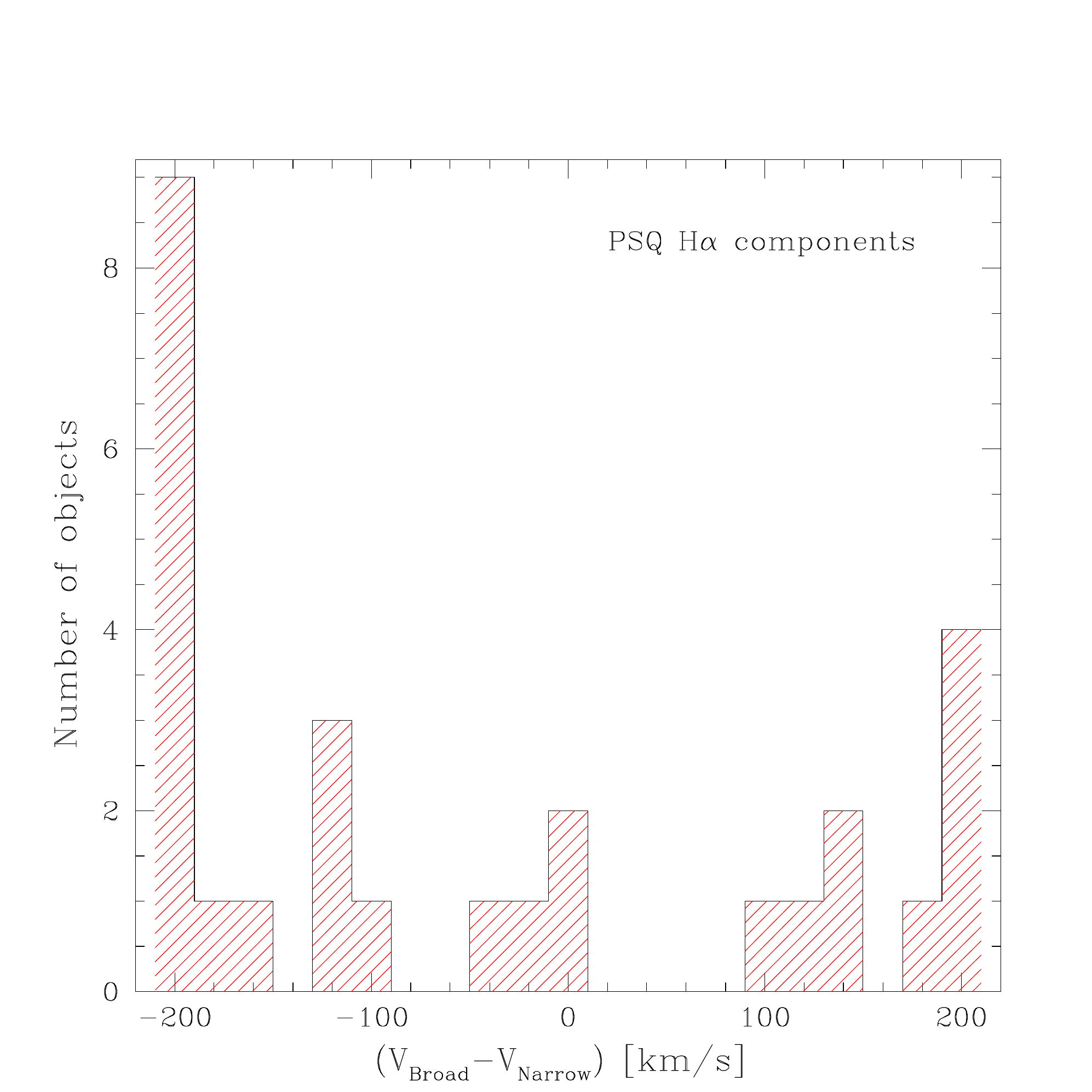}\includegraphics[height=7cm]{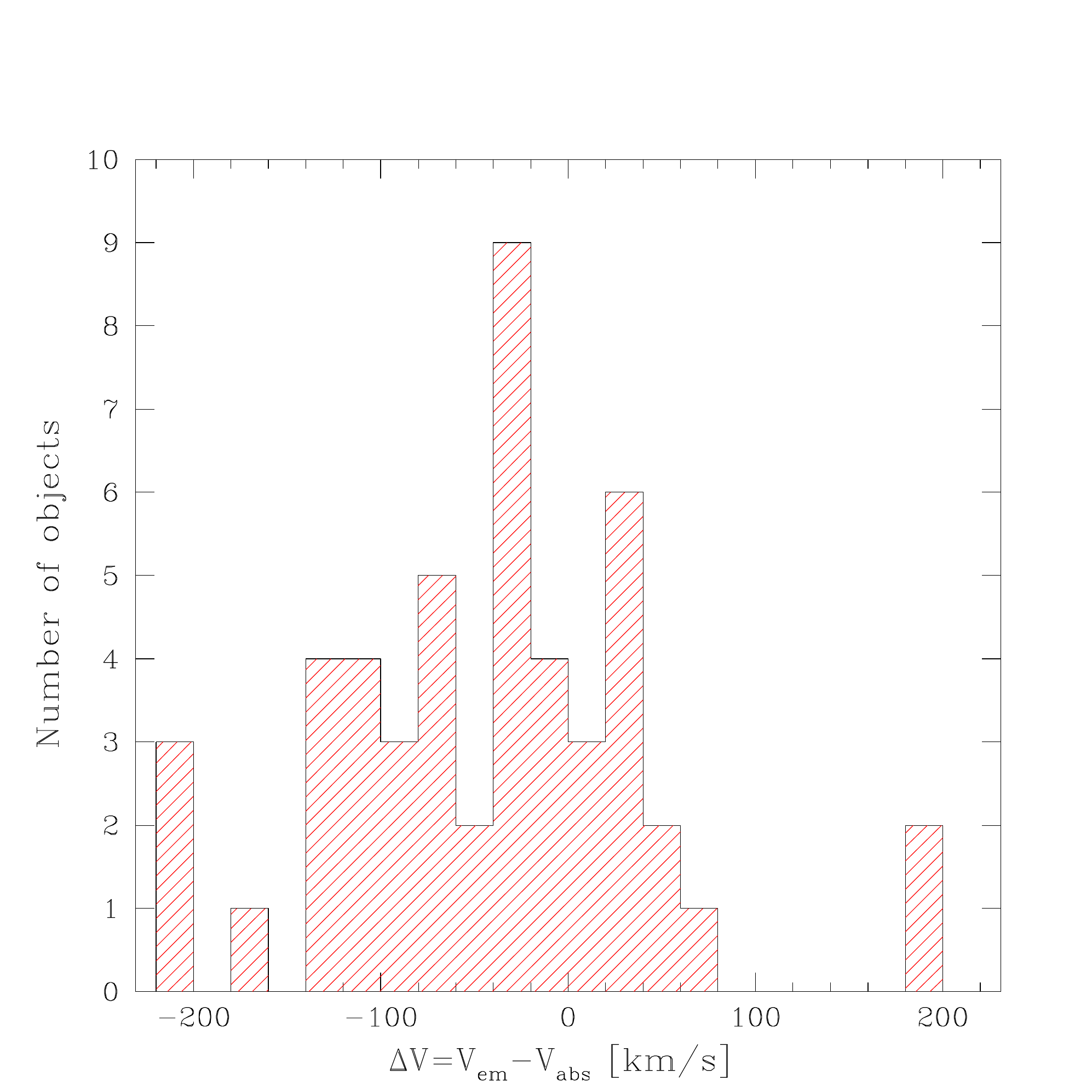}
\caption {{\bf Left} Radial velocity differences between the broad and narrow components of the H$\alpha$ line measured fitting multiple Gaussians.  
{\bf Right}. Histogram of the difference in radial velocity between the emission lines ($V_{em}$ and the absorption lines ($V_{abs}$). The measurement errors are about $\pm$20km/s.}
 \label{velas}
\end{figure*}

In the process of separating the broad and narrow components of H$\alpha$ we noticed that for the majority of our objects the broad component appeared to be systematically shifted 
relative to the narrow components, mostly, but not in all cases, to the blue.  This is illustrated in the left panel of Figure~\ref{velas} that plots the radial velocity 
difference between the centroids of the broad and narrow lines measured from the multiple Gaussian fits. 

The right-panel of the figure shows the velocity shift between the narrow components (in most cases measured from the two [OIII]4959,5007 lines) and the stellar lines as measured by {\sc Starlight} from cross-correlation with the best fitting stellar templates.  Negative values correspond to gas moving towards the observer and positive values to gas moving away from the
observer relative to the stars.   From repeated fits and from the two [OIII] emission lines we estimate the (star-gas) radial velocity differences to be accurate to $\pm20$km/s, so the figure shows a clear excess of blue shifted gas emission.  

\begin{figure*}
\includegraphics[height=7cm,angle=-90]{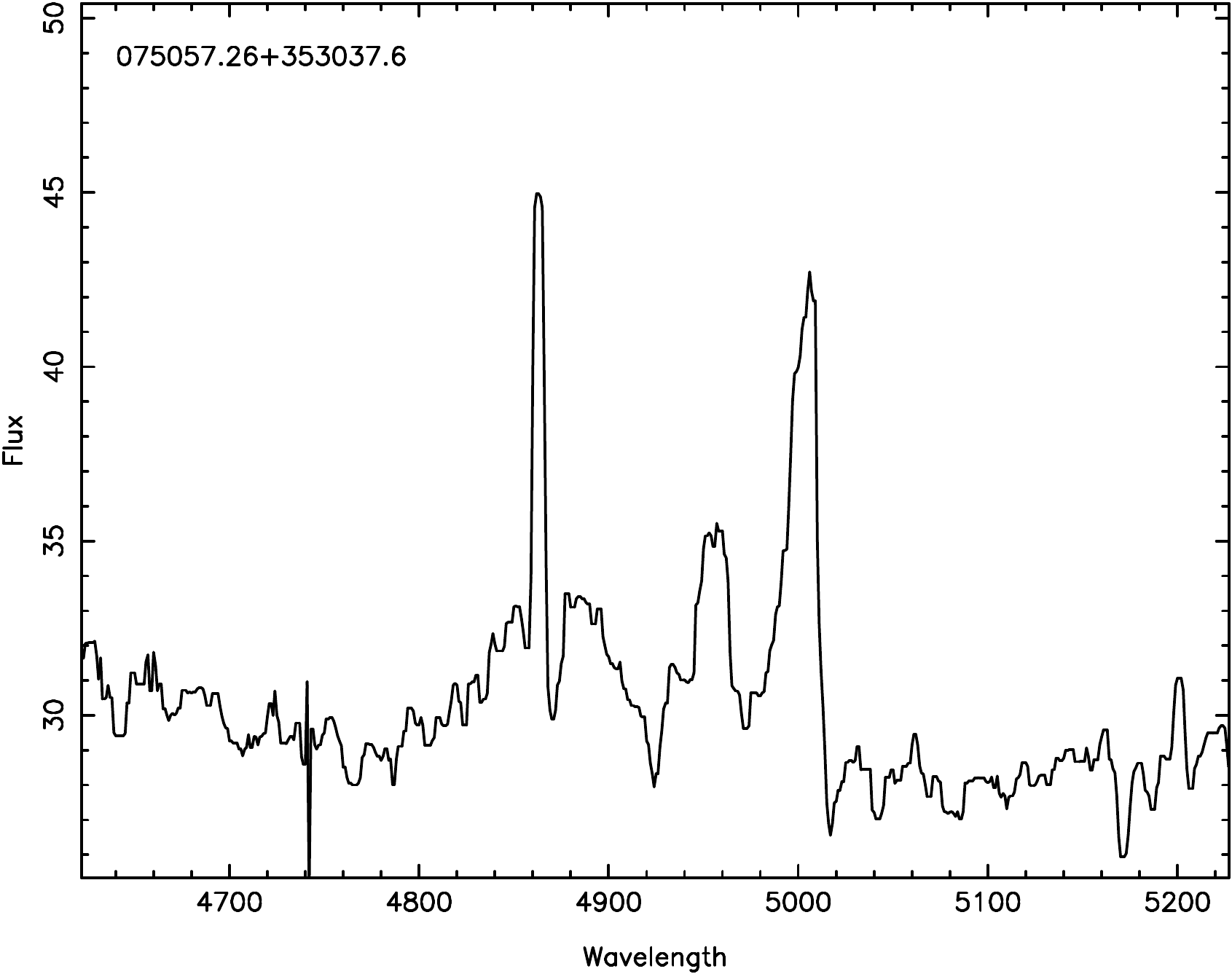}\hspace{1cm}\includegraphics[height=7cm,angle=-90]{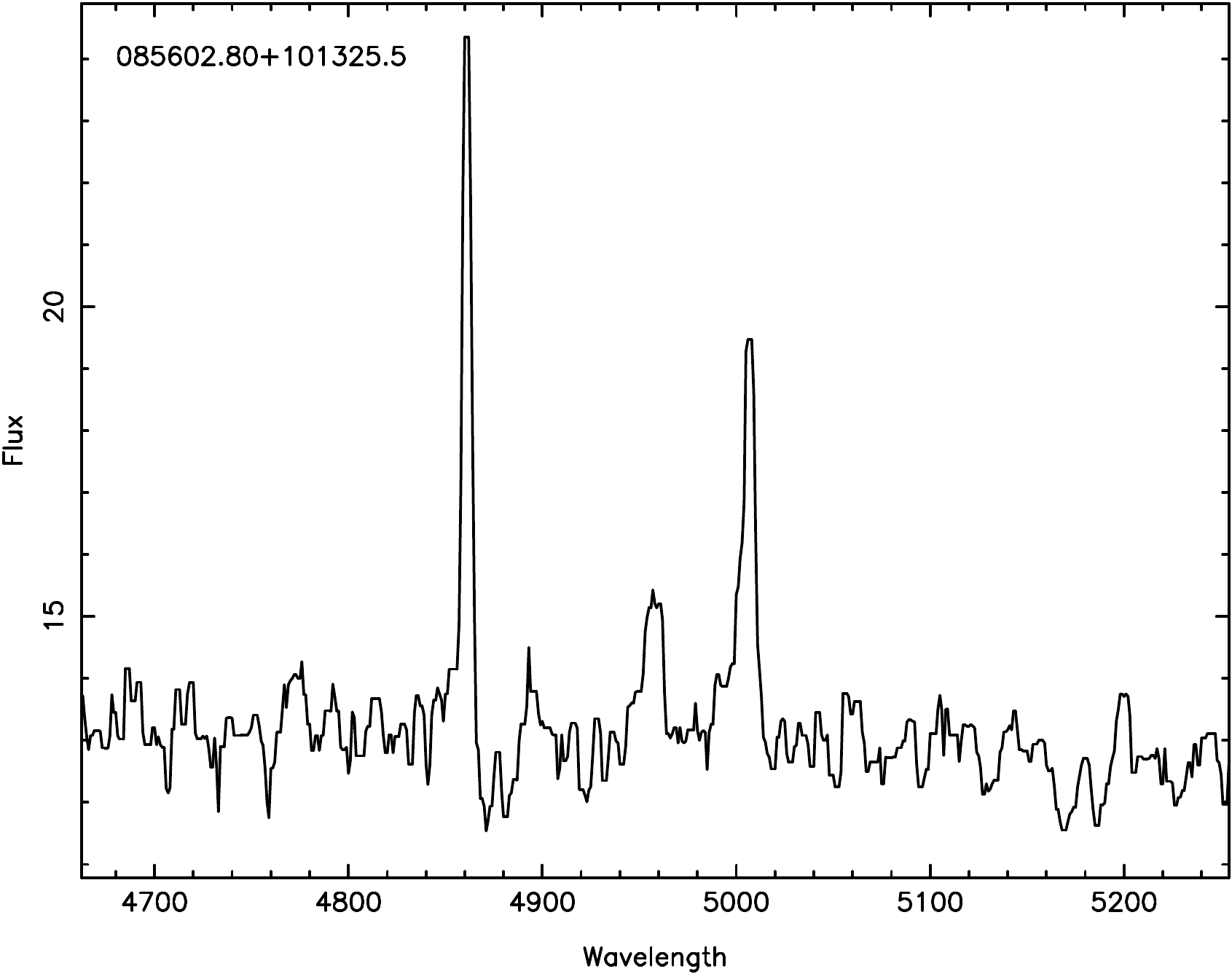}
\includegraphics[height=7cm,angle=-90]{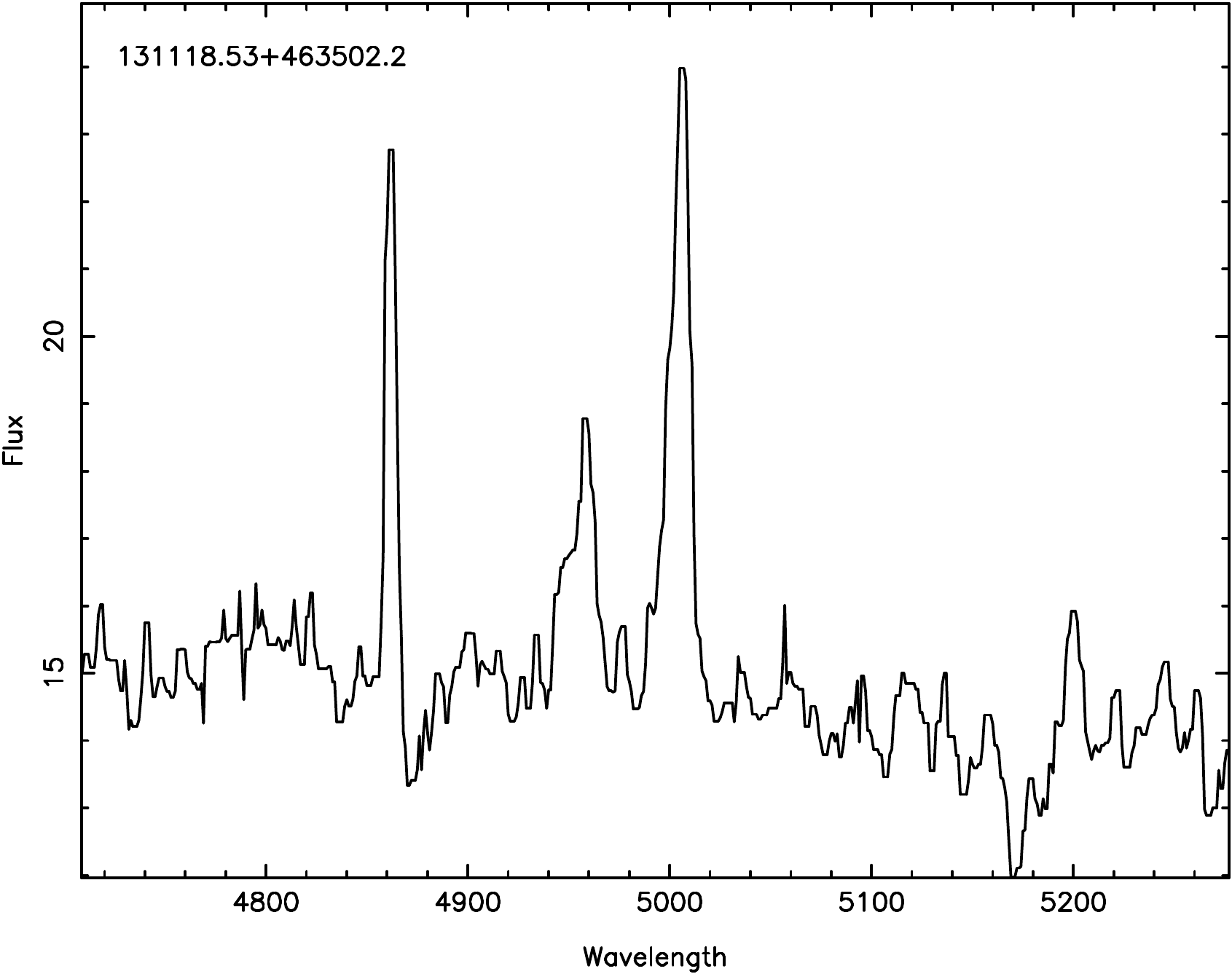}\hspace{1cm}\includegraphics[height=7cm,angle=-90]{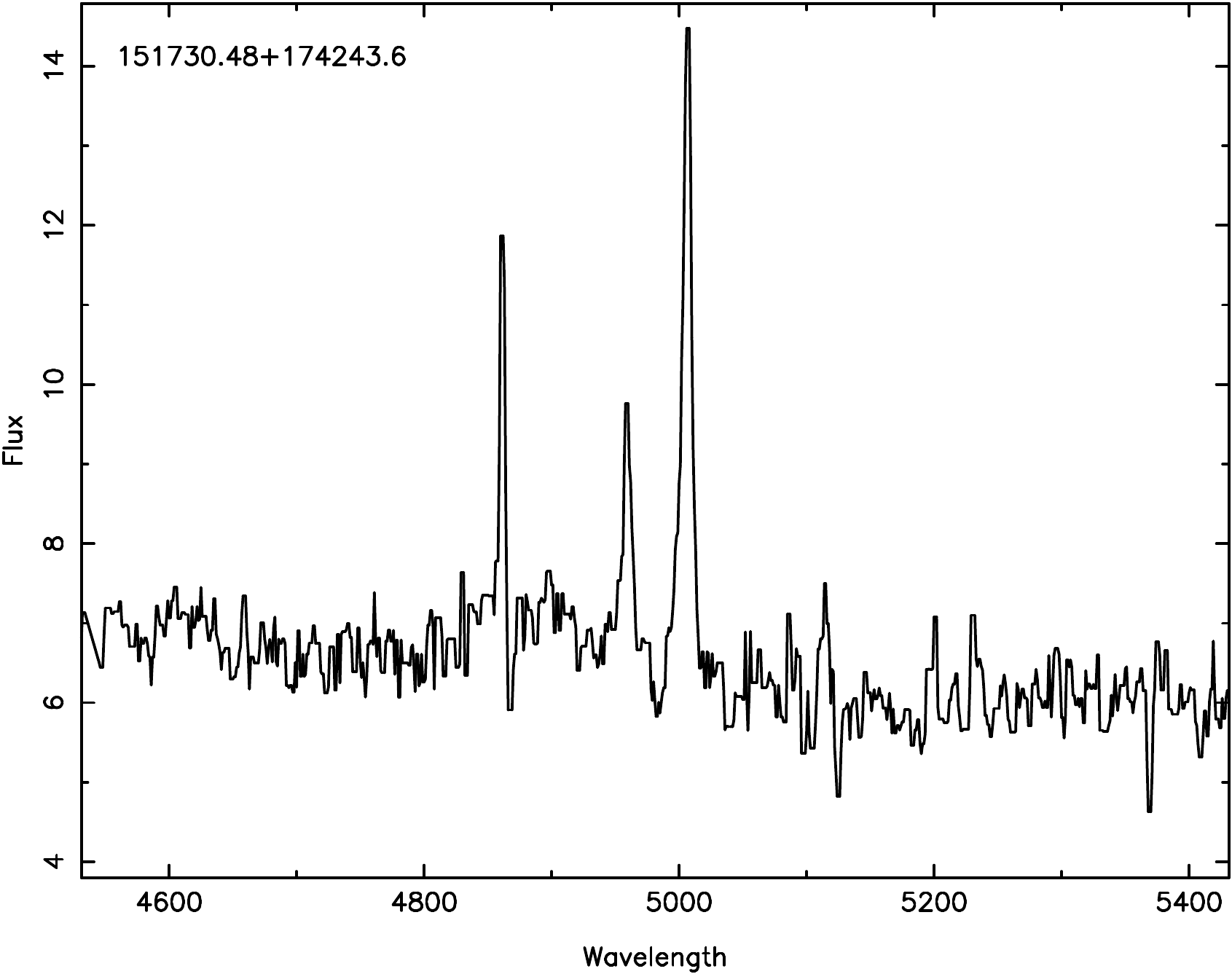}
\caption {Examples of some of the few objects in our sample that have narrow and relatively strong H$\beta$ emission lines. The P-Cygni profiles show that the gas is blue shifted relative to the stellar component. }
 \label{hbeta}
\end{figure*}

In general the H$\beta$ emission line in PSQ is either completely absorbed by the underlying absorption, or broad and rather weak, but for some cases H$\beta$ is strong and narrow showing clear P-Cygni profiles. Some of these cases are shown Figure~\ref{hbeta} that show blue-shifted emission relative to the stellar line, confirming the predominance of blue shifts for the emission lines. 


\subsection{Black-Hole masses and X-ray luminosity}

{ The simplest way to determine  Black-Hole masses for our PSQs is either from the Eddington luminosities, or from the masses of the spheroidal component of the host galaxies. The statistical study of X-ray luminous AGN by \cite{Wu2012} provides a useful way to estimate the Eddington ratio $\eta_{Edd}=L_{Bol}/L_{Edd}$ from the slope of the power-law continuum in the UV. The average power-law slope for our sample $\alpha_{UV}=-1.5$  (in $f_{\lambda}$ units)  is typical for the QSOs in the sample of \cite{Wu2012} and indicates an Eddington ratio $\eta_{Edd}\sim0.3$.

\cite{Runnoe2012} provide bolometric luminosity corrections for quasars at $0.145\mu$ and $0.3\mu$. Assuming a power-law slope $\alpha_{UV}=-1.5$ for all the objects in our sample, we can use the dereddened power-law luminosities at these wavelengths to estimate the bolometric luminosities using the recommended relations of \cite{Runnoe2012}. We then use the Eddington luminosity,  $L_{Edd}=3\times10^4M_{BH}$ in solar units, to compute Black-Hole masses M$^{Edd}_{BH}$ as

\begin{equation}
M^{Edd}_{BH} = \frac{L_{Bol}}{3\times10^4\eta_{Edd}}M_{\odot},
\end{equation}
\smallskip\\
where as before $\eta_{Edd}$ is the Eddington ratio. The resulting distribution of black-hole masses is shown by the black histogram in Figure~\ref{mbh}.

The second histogram in red shows the BH masses estimated from the mass of the spheroidal component $M_{sph}$ using the total stellar masses $M_*$ from the population synthesis models as a proxy for $M_{sph}$. For simplicity we assumed $M_{sph}=0.85M_*$ for all our objects, and used the correlation between $M_{sph}$ and $M_{BH}$ from \cite{Scott2013}:  $logM^*_{BH}=0.97log(M_{sph}/3\times10^{11})+9.27$  to infer BH masses. 

\begin{figure}
\includegraphics[height=8cm]{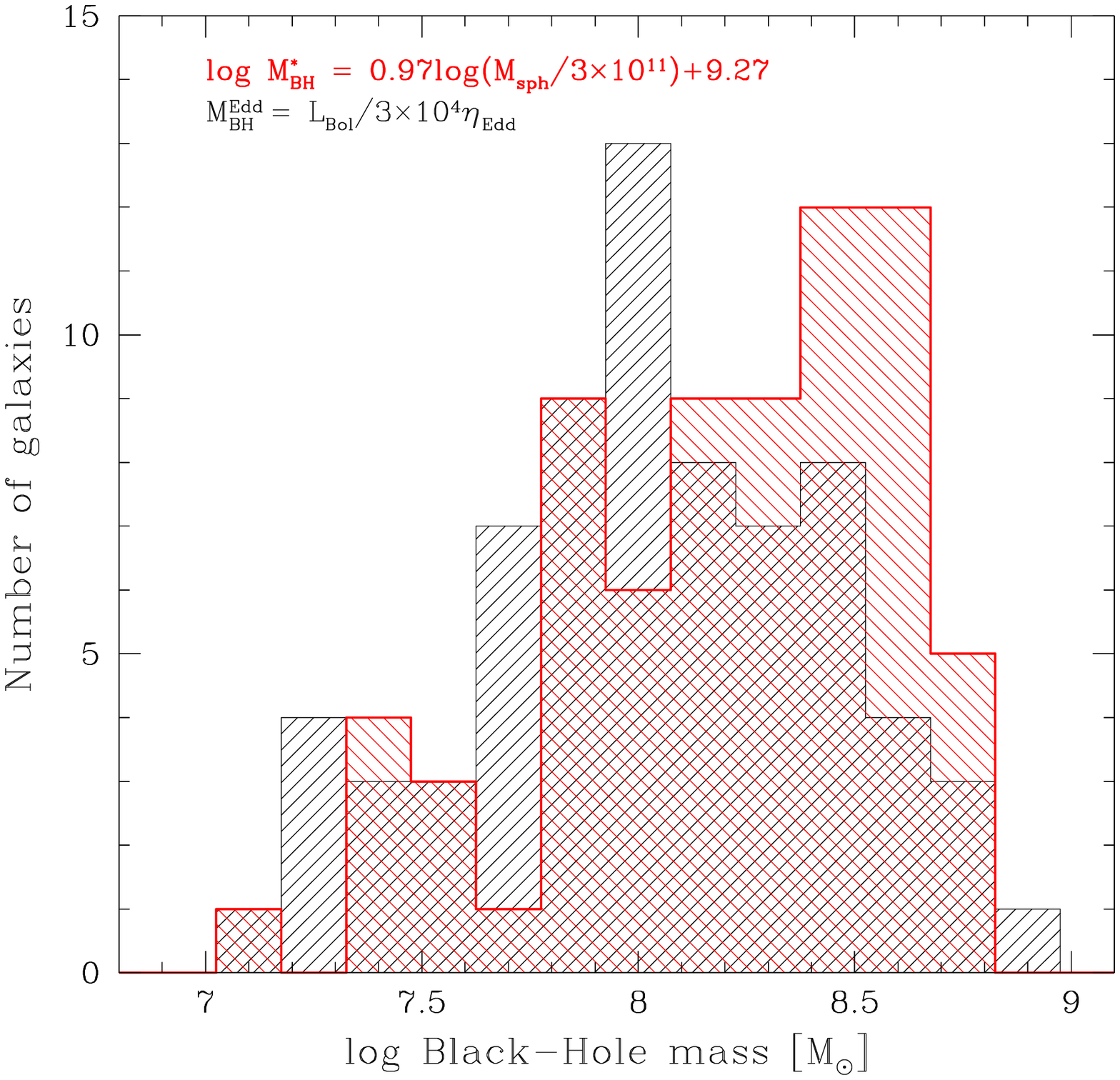} 
\caption { {\bf left}. Black-Hole masses calculated with two different methods shown in the lagend. The histogram in black (lleft) was calculated using the Eddington ratios and the bolometric luminosities. The histogram in red (right) shows the distribution of masses calculated using the relation between $M_{BH}$ and $M_{sph}$. {\bf right}. Correlation between $M^{Edd}_{BH}$ and $M^*_{BH}$. The solid line shows a least-squares fit with coefficients indicated in the legend.  }
 \label{mbh}
\end{figure}

{Two distributions in Figure~\ref{mbh}  are in remarkably good agreement considering the approximations that we have made, in particular using the same power-law and the same stellar mass to bulge mass ratios for all our objects. Still,  the Eddington ratio $\eta_{Edd}=0.3$  depends only weakly on the UV power-law slope \citep{Wu2012}, while if anything the assumption of $M_{sph}=0.85M_*$ is rather conservative. 

The largest uncertainty arises because a substantial fraction of the total luminosity at $0.55\mu$ is contributed by the power-law component (Fig.~\ref{seds}). So, in order to calculate the total stellar masses of each object, we have assumed that for all the objects in our sample the power-law component contributes 55\% of the visible light. This correspond to the ratio of stellar to power-law luminosity of $L_*/L_{PL}=0.81$ (Figure~\ref{seds}).

\cite{Cales2013} studied a sample of 38 PSQ with substantially better S/N spectra, and using the more standard method of combining line widths and optical luminosities. Their masses for 12 objects in common are comparable, within the uncertainties, with our $M^*_{BH}$ values, indicating that the assumptions that went into our calculations are not entirely unreasonable.}

An independent test is to use the X-ray luminosities derived from the bolometric corrections. \cite{Runnoe2012} give two separate relations for radio-loud and radio-quiet objects. All the PSQ in our sample are in the NVSS \citep{Condon1998} and, as expected, the vast majority are radio quiet. Only six objects in our sample are radio loud with fluxes well in excess of 3mJy/beam, which seems to be consistent with the fact that at similar luminosities only about 10\% of all QSOs are radio loud \citep{Padovani1993}.  Inverting the bolometric correction for radio-quiet QSOs: $logL_{Bol}=33.08+0.29logL_{2-10keV}$ we get the distribution of 2-10keV X-ray luminosities shown in Figure~\ref{lx}.

\begin{figure}
\includegraphics[height=8cm]{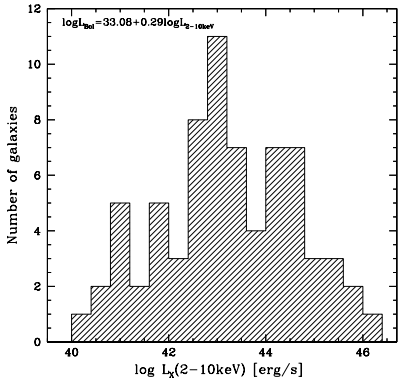}
\caption {Distribution of X-ray luminosities for PSQ calculated using the bolometric corrections for radio-quiet QSOs shown in the legend. }
 \label{lx}
\end{figure}

{ At these luminosities we would expect only three of our objects to be detected by ROSAT, while none appears in the ROSAT all sky catalogue \citep{Voges1999}, which at the mean redshift of our sample has a $3\sigma$ detection limit of $logL_{2-10keV}\sim44.3$ erg/s. Nine of our PSQ are in the XMM or Chandra archives, but only two (J025938.15+004206.4 and J074524.97+375436.6) are robust detections, and a third (J094820.38+582526.4) is an upper limit. The mean X-ray luminosity of these objects, $log L_X=43.2$ erg/s, is consistent on average with our estimates from the bolometric luminosities ($log L_X=43.1$). We can confidently conclude, therefore,  that the AGN in PSQ appear to be normal from the point of view of their black-hole masses, and their radio and X-ray luminosities.}



\section{Discussion}
\subsection{Kinematics}

In the scenario of \cite{Hopkins08a}, the QSO phase is preceded by the Antennae/ULIRG phases characterised by the formation of numerous starburst clusters and super-clusters in the arms of both spiral components. While large amounts of molecular material may be channelled to the nuclear regions by tidal forces during the merger, nuclear starbursts in these phases are either still weak, or heavily enshrouded in dust such that their emission is not visible at optical and near-IR wavelengths.  By the time massive starburst activity is finally ignited in the nucleus, all star formation in the arms would have ceased leaving behind a large intermediate age population of stars distributed across the remains of the tidally distorted spiral arms.

Therefore, while the emission lines in PSQ mostly arise from the nuclear regions (e.g. \citealt{Storchi2014}), the stellar lines originate mostly from the remnants of massive starburst clusters in the spiral arms, so it should come as no surprise to observe a wide distribution of radial velocity shifts between stars and gas. And, since these shifts depend mostly on the  { --  in principle random -- orientation of the merger with respect to the line of sight}, we would expect to observe a uniform distribution of velocity-shifts centred around zero. However, as shown in  the right panel of Figure~\ref{velas}, while we do observe a uniform distribution, the gas appears to be systematically blue shifted relative to the stars.

On the other hand, we observe a rather messy distribution of relative velocities between the narrow and broad components of $H\alpha$, with the broad components either blue-shifted or red-shifted relative to the narrow lines.   { Therefore, if the broad-line emitting region (BLR) is at rest relative to the systemic velocity of the galaxy, the centroid shifts we observe must be due to a combination of extinction and inflow, outflow, or rotation of the BLR.}

For a spherical geometry, the predominance of blue shifts in our sample would indicate a net outflow of the BLR, contrary to the model of  \cite{Gaskell2013}, but other geometries, such as an inclined rotating disk plus an oblique obscuring torus, could also provide the variety of blue and red shifts observed in PSQ \citep{Stern2015}. Beyond asserting that there are ways to explain the observations of PSQ, however, the kinematics of the BLR is a very complex issue that is well beyond the scope of this paper. 
 
Nevertheless, we  may safely conclude that the distribution of the relative velocities between the narrow emission lines and the stars result from a combination of rotation plus a net outflow (in most cases) of the narrow emission line emitting gas. This conclusion agrees with the spatially resolved observations of the nearby PSQ J0330-0532 by \cite{Storchi2014}, although this object is not in our catalogue as it hosts a rather small population of intermediate-age stars.

\subsection{Nuclear star formation}
We have seen that the intermediate-age stellar populations observed in K+A galaxies are not in place in PSQ, where intermediate-age stars are both less abundant and significantly metal poorer. We are forced to conclude, therefore, that either PSQ are not the immediate progenitors of K+A galaxies, or that the bulk of the young stars in PSQ are still being formed. This requires that vast amounts of metal-enriched molecular gas be funnelled onto the central regions during the merger in order to fuel a huge nuclear starburst, which is still not visible in PSQ.
Unfortunately, the SDSS spectra of our objects lack the S/N and resolution required to infer the metallicities of the nebular gas in PSQ and discern nuclear stellar features, but we do know that in K+As the little nebular gas that remains has well over solar metallicities.

The star formation rates of these putative nuclear starbursts would need to reach 10-100 solar masses per year ($\sim10^{10}M_{\odot}$  in less than a few 100Myr) and be extremely compact, perhaps extending over not much more than a few kpc \citep{Storchi2014}. Under these conditions it would seem reasonable to expect that much of the luminosity that we observe as an obscured power-law component in PSQ would be produced not by an AGN, but by a very massive nuclear starburst. At an age of a few Myr the bolometric mass-to-light ratio of a simple starburst is about $M/L_{Bol}\sim3\times10^{-4}$ \citep{Leitherer99}, so the mean bolometric luminosity of the power-law component in our PSQ, $L_{Bol}=10^{44.9} erg~s^{-1}$, would imply a total mass of about $10^{8}M_{\odot}$ in young stars, about two orders of magnitude lower than the "missing" intermediate-age stellar population in PSQ.  

An additional challenge for this scenario is that the young stars must form over time scales much shorter than 10Myr, which requires a very compact concentration of molecular gas. Compact nuclear  starburst clusters may resemble type-I AGN as a consequence of supernovae evolving in a high density interstellar medium \citep{Terlevich92}, so a critical test of this Starburst-AGN connection is that at least some PSQ should have huge concentrations of cold, metal-rich molecular gas in their nuclear regions making them extremely luminous CO sources.
 
\subsection{Comparison with previous investigations}

{ \cite{Canalizo2001} have studied a sample of {\em transitional} AGN which seem to be objects in transition between the (U)LIRG and the PSQ phase. In \cite{Canalizo2013} they extended their study to more normal QSOs and consistently find that the intermediate-age populations of their objects - that they associate with remnants of major mergers - are offset from the nuclei by typically $\sim8kpc$. This is consistent with our conjecture that the intermediate-age populations in PSQ must be the remnants of starburst clusters in the arms of the merging galaxies.}

Our results are in broad agreement with the study of \cite{Cales2013} who observed 38 PSQ selected from the SDSS, in particular with our conclusion that the putative AGN in PSQ coexist with massive bursts of star formation. Unfortunately, there are only 12 objects in common between our two samples. Ten of the objects in the \cite{Cales2013} sample are not in the DR7Q, while their selection criterion EW(H$\delta)>$1\AA\ is substantially less stringent than our criterion of EW(H$\delta)>$3\AA, so it admits objects that we would not classify as PSQ. 

{ While as expected the majority of our objects fall in the strong AGN (sAGN) domain of the WHAN diagnostic diagram, most of the objects in the Cales et al. sample appear to be either LINERS or star-forming, with the caveat that they use the [OIII]/H$\beta$ vs. [NII]/H$\alpha$ BPT plane to classify their objects. In PSQ H$\beta$ is rather weak and extremely sensitive to corrections for underlying absorption, so using BPT diagrams involving H$\beta$ seems risky. However, the BH masses derived by \cite{Cales2013} using a completely different methodology are substantially similar to the BH masses of our PSQ, and this indicates that, even thought the stellar populations may differ, the AGNs themselves seem to be similar.}

In their study of J0330-0532, \cite{Storchi2014} found that the intermediate-age population avoids the nuclear regions and is distributed along a disk-like structure, which would be consistent with our results, although as mentioned above, this object does not meet our selection criterion for PSQ.  

\section{Conclusions}

We have compiled a sample of 72 {bona fide} post-starburst QSOs (PSQ) selected from the SDSS DR7 quasar catalogue purely on the basis of the equivalent width of the Balmer H$\delta$  lines.  We have used published photometry and SDSS spectroscopy to study the stellar populations, gas properties, and SEDs of the sample. Our main conclusion is that PSQ do not (yet) contain the large population of metal-rich intermediate-age stars that we observe in post-starburst (K+A) galaxies. Thus, either PSQ are not the immediate progenitors of K+As, or they contain hugely massive nuclear starbursts where the bulk of their intermediate-age stars are still being formed.

Our results, however, are based on rather low S/N SDSS spectra of these relatively faint and distant objects. Further progress in the understanding of PSQ, therefore, requires substantially deeper observations in order to refine the population synthesis models and to investigate the kinematics and physical conditions of the nebular gas as well as to peek into the nuclear regions in search for the hidden AGN. 

The majority of our PSQ are not strong radio or X-ray sources, although the number of objects observed in X-rays remains rather small. Thus, deep X-ray and mm wave observations are critically needed to establish the nature of the AGN in these objects and to check for the presence of nuclear starbursts. The present paper reports the initial steps of such research programme.

\section*{Acknowledgements}
JM acknowledges the award of Special Visiting Researcher fellowship of the {\em Ciencia sem fronteiras} programme of the Brazilian government through their federal funding agencies - CNPq. JM also thanks the hospitality of Nanjing University where the initial steps of this research programme were outlined.

We are grateful to Roberto Cid-Fernandes the father of {\sc Starlight} for continuous support in the use of the code. The {\sc Starlight} project is supported by the Brazilian agencies CNPq, CAPES and FAPES and by the France-Brazil CAPES/Cofecub programme.

We thank our anonymous referee for very useful comments that led to a much improved version of the paper.
 
This research has made use of the NASA/ IPAC Infrared Science Archive, which is operated by 
the Jet Propulsion Laboratory, California Institute of Technology, under contract with the 
National Aeronautics and Space Administration. 


The entire GALEX Team gratefully acknowledges NASA's support for construction, operation, and science 
analysis for the GALEX mission, developed in corporation with the Centre National d'Etudes Spatiales of 
France and the Korean Ministry of Science and Technology. We acknowledge the dedicated team of engineers,
technicians, and administrative staff from JPL/Caltech, Orbital Sciences Corporation, University of California, 
Berkeley, Laboratoire d'Astrophysique Marseille, and the other institutions who made this mission possible. 

Funding for the SDSS and SDSS-II has been provided by the Alfred P. Sloan Foundation, the Participating 
Institutions, the National Science Foundation, the U.S. Department of Energy, the National Aeronautics and 
Space Administration, the Japanese Monbukagakusho, the Max Planck Society, and the Higher Education
Funding Council for England. The SDSS Web Site is http://www.sdss.org/.

The SDSS is managed by the Astrophysical Research Consortium for the Participating Institutions. The
Participating Institutions are the American Museum of Natural History, Astrophysical Institute Potsdam, 
University of Basel, University of Cambridge, Case Western Reserve University, University of Chicago, 
Drexel University, Fermilab, the Institute for Advanced Study, the Japan Participation Group, Johns 
Hopkins University, the Joint Institute for Nuclear Astrophysics, the Kavli Institute for Particle Astrophysics 
and Cosmology, the Korean Scientist Group, the Chinese Academy of Sciences (LAMOST), Los Alamos 
National Laboratory, the Max-Planck-Institute for Astronomy (MPIA), the Max-Planck-Institute for Astrophysics 
(MPA), New Mexico State University, Ohio State University, University of Pittsburgh, University of Portsmouth,
Princeton University, the United States Naval Observatory, and the University of Washington.
 
This publication makes use of data products from the Two Micron All Sky Survey, which is a joint project of 
the University of Massachusetts and the Infrared Processing and Analysis Center/California Institute of 
Technology, funded by the National Aeronautics and Space Administration and the National Science 
Foundation


This publication makes use of data products from the Wide-field Infrared Survey 
Explorer, which is a joint project of the University of California, Los Angeles, and 
the Jet Propulsion Laboratory/California Institute of Technology, funded by the 
National Aeronautics and Space Administration.

This research has made use of the NASA/IPAC Extragalactic Database (NED) 
which is operated by the Jet Propulsion Laboratory, California Institute of Technology, 
under contract with the National Aeronautics and Space Administration.


 
\end{document}